\documentstyle[12pt]{article}
\begin{document}
\begin{center}
{\large\bf What's Wrong with Pauli-Villars Regularization}

\vspace{0.2cm}

{\large\bf in $QED_{3}$ ?}
\vspace{1.0cm}\\
B. M. Pimentel\footnote{Partially supported by CNPq} and J. L.
Tomazelli\footnote{Supported by CAPES}
\vspace{0.2cm}

Instituto de F\'{\i}sica Te\'{o}rica \\
Universidade Estadual Paulista \\
Rua Pamplona, 145 \\
01405--900 - S\~{a}o Paulo, SP - Brazil

\vspace{1.0cm}

{\bf Abstract}
\end{center}
\vspace{0.3cm}

In this letter we argue that there is no ambiguity between the Pauli-Villars
and other methods of regularization in (2+1)-dimensional quantum
electrodynamics with respect to dynamical mass generation, provided we
properly choose the couplings for the regulators.

\vspace{0.4cm}

electromagnetic interactions.

\vspace{1.0cm}

It is well known that gauge theories in (2+1)-dimensional
space-time${}^{[1]}$, though super-renormalizable, show up inconsistencies
already at one loop, arising from the regularization procedure adopted to
evaluate ultraviolet divergent amplitudes such as the photon self-energy in
spinor quantum electrodynamics. In the latter, if we use analytic${}^{[2]}$
or dimensional${}^{[3]}$ regularization, the photon is induced a topological
mass, in contrast with the result obtained through the
Pauli-Villars${}^{[1]}$ scheme, where the photon remains massless when we
let the auxiliary masses go to infinity.

Recently, alternative constructs${}^{[4],[5]}$ have been proposed in order to
cla\-rify this matter, which essentially rely upon  using causal dispersion
relations. They put forward that the photon indeed dynamically acquires a
topological mass. Thus, we are led to ask whether the ordinary Pauli-Villars
prescription shouldn't be taken with a grain of salt.

Let us begin by close examining the conditions that must be imposed on the
masses and coupling constants of the regulator fields such that a
regularized closed fermion loop in 2+1 dimensions is rendered finite along the
calculations. Consider the integral corresponding to a fermion loop with $n$
vertices to which we associate $n$ external photon lines with momenta $k_i$
($i$=1,2,...,$n$). This integral is proportional to
\begin{equation}
\int d^3p\, \frac{Tr[\gamma_{\mu_1}(m+\slash\!\!\!p)\gamma_{\mu_2}
(m+\slash\!\!\!p+\slash\!\!\!k_1)...\gamma_{\mu_n}(m+\slash\!\!\!p+...+
\slash\!\!\!k_{n-1})]}{(m^2-p^2+i\epsilon)[m^2-(p+k_1)^2+i\epsilon]...
[m^2-(p+...+k_{n-1})^2+i\epsilon]}
\end{equation}
so, for large $p$, its integrand behaves like $p^{-n}$ whereas for $n<4$ the
integral diverges as

\[\int_0^{\infty} \,\frac{p^2\,dp}{p^n} \sim \int_0^{\infty}\frac{dp}
{p^{n-2}}\,\,.\]

This integrand can be written as
\begin{equation}
{\cal I} \equiv \frac{P_n(p)+mP_{n-1}(p)+m^2P_{n-2}(p)+...+m^n}{P_{2n}(p)
+m^2P_{2n-2}(p)+...+m^{2n}}\,\,,
\end{equation}
where $P_i(p)$ stands for a polynomial of degree $i$ in the components of
$p$. We can write the denominator of ${\cal I}$ in the form
\[P_{2n}(p)\left(1+m^2\frac{P_{2n-2}(p)}{P_{2n}(p)}+...+m^{2n}\frac{1}
{P_{2n}(p)}\right)\]
and, for large $p$, perform the expansion
\[\frac{1}{\left(1+m^2\frac{P_{2n-2}(p)}{P_{2n}(p)}+...+
m^{2n}\frac{1}{P_{2n}(p)}\right)} \sim 1-m^2\frac{P_{2n-2}(p)}{P_{2n}(p)}
+...\,\,,\]
so that the integrand ${\cal I}$ behaves like
\[{\cal I} \sim \frac{P_{n}}{P_{2n}}-m^2\frac{P_{n}P_{2n-2}}{P_{2n}P_{2n}}+
m\frac{P_{n-1}}{P_{2n}}-m^3\frac{P_{n-1}P_{2n-2}}{P_{2n}P_{2n}}+
m^2\frac{P_{n-2}}{P_{2n}}+...\,\,,\]
i.e.,
\begin{eqnarray}
{\cal I} & \sim & \displaystyle{\frac{P_{n}}{P_{2n}}+m\frac{P_{n-1}}
{P_{2n}}+m^2\frac{P_{n}}{P_{2n}}
\left[\frac{P_{n-2}}{P_{n}}-\frac{P_{2n-2}}{P_{2n}}\right]+} \nonumber \\
         &      & \displaystyle{m^3\frac{P_{n-1}}{P_{2n}}\left[\frac{P_{n-3}}
{P_{n-1}}-\frac{P_{2n-2}}{P_{2n}}\right]+...} \nonumber \\
         &  =   & \displaystyle{\sum_{k} m^k\,a_{-(n+k)}(p)\,\,,}
\end{eqnarray}
where
\[a_{-(n+k)}(p) \sim p^{-(n+k)}\,\,. \]
Therefore, in making the substitution
\[{\cal I}(m) \rightarrow \sum_{i}^{n_f} c_{i}{\cal I}(M_i)\,\,,\]
where $n_f$ is the number of auxiliary fermion fields, we must impose in the
vacuum polarization case ($n=2$) the following conditions:
\begin{eqnarray}
\sum_{i}^{n_f} c_{i}=0\,\,, \\
\sum_{i}^{n_f} c_{i}M_i=0\,\,,
\end{eqnarray}
in order to eliminate the linear and logarithmic divergences, respectively.

Having settled down the basis for the Pauli-Villars regularization method, we
turn to the calculation of the vacuum polarization tensor in spinor $QED_3$.
In the standard notation, the regularized expression for the vacuum
pola\-rization tensor reads
\begin{equation}
\Pi^{M}_{\mu \nu}(k)=\frac{ie^2}{(2\pi)^3}\sum_{i=0}^{n_{f}}c_{i}\int d^3p
\frac{P(M_{i})}{({M_{i}}^2-{p_{1}}^2)({M_{i}}^2-{p_{2}}^2)} \,\,,
\end{equation}
where
\begin{eqnarray}
&c_{o}=1
\hspace{0.3cm},
\hspace{0.3cm}
M_{0}=m
\hspace{0.3cm},
\hspace{0.3cm}
M_i=m{\lambda}_i\,\,(i=1,...,n_f) \,\,\,, \\
&p_{1,2}=p \mp \frac{1}{2}k
\end{eqnarray}
and
\begin{eqnarray}
&P(M_{i})= Tr \{\gamma_{\mu}(p_{1}\!\!\!\!\!/ \,+M_{i})
\gamma_{\nu}(p_{2}\!\!\!\!\!/ \,+M_{i})\} \nonumber \\
&=2[{M_{i}}^2 g_{\mu \nu}+{p_{1}}_{\mu}{p_{2}}_{\nu}+{p_{1}}_{\nu}{p_{2}}_{\mu}
-g_{\mu \nu}(p_{1}.p_{2})-iM_{i}\epsilon_{\mu \nu \alpha}k^{\alpha}] \,\,.
\end{eqnarray}
For simplicity, but without loss of generality, we may choose both the
electron mass and those of the auxiliary fields to be positive; the
coefficients ${\lambda}_i$ ultimately go to infinity to recover the original
theory. Using the Feynman parametrization
\begin{equation}
\frac{1}{({M_{i}}^2-{p_{1}}^2)({M_{i}}^2-{p_{2}}^2)}=
\int_{0}^{1} d\,\xi \frac{1}{{[{M_{i}}^2-{p_{1}}^2
-({p_{2}}^2-{p_{1}}^2)\xi]}^2}
\end{equation}
and performing the momentum shift $p_{\mu} \rightarrow
p_{\mu}+( \frac{1}{2}-\xi)k_{\mu}$, we get
\begin{equation}
\Pi^{M}_{\mu \nu}(k)=(g_{\mu \nu}-\frac{k_{\mu}k_{\nu}}{k^2})
\Pi^{M}_{1}(k^2)
+im\epsilon_{\mu \nu \alpha}k^{\alpha}\Pi^{M}_{2}(k^2)+
\Pi^{M}_{GB}(k^2)\,\,,
\end{equation}
where
\begin{eqnarray}
\Pi^{M}_{1}(k^2) & \equiv & \displaystyle{4ie^2k^2 \sum_{i=0}^{n_{f}}c_{i}
\int_{0}^{1}d\,\xi\,\,\xi(1-\xi) \int \frac{d^3p}{{(2\pi)}^3} \frac{1}
{{({Q_{i}}^2-p^2)}^2}} \,\,, \\
\Pi^{M}_{2}(k^2) & \equiv & \displaystyle{-\frac{2ie^2}{m} \sum_{i=0}^{n_{f}}
c_{i}M_{i}\int_{0}^{1} d\,\xi \int \frac{d^3p}{{(2\pi)}^3}
\frac{1}{{({Q_{i}}^2-p^2)}^2}} \,\,, \\
\Pi^{M}_{GB}(k^2)  & \equiv & \displaystyle{\frac{2}{3}ie^2g_{\mu \nu}
\sum_{i=0}^{n_{f}}c_{i}\{I^{1}_{i}+I^{2}_{i}\}} \,\,,
\end{eqnarray}
with the following defininitions
\begin{eqnarray}
I^{1}_{i} \equiv 3\int_{0}^{1} d\,\xi \int \frac{d^3p}{{(2\pi)}^3}
\frac{1}{({Q_{i}}^2-p^2)} \,\,, \\
I^{2}_{i} \equiv 2\int_{0}^{1} d\,\xi \int \frac{d^3p}{{(2\pi)}^3}
\frac{p^2}{{({Q_{i}}^2-p^2)}^2}
\end{eqnarray}
and
\begin{equation}
Q_{i}^2 \equiv M_{i}^2-\xi\,(1-\xi)k^2 \,\,.
\end{equation}

If we carry out the momentum integrations in (15) and (16) it is
straightforward to arrive at $\Pi^M_{GB}(k^2) \equiv 0$, as expected by gauge
invariance. Here comes the crucial point: we can't blindly take only one
auxiliary field with $M=\lambda m$ as usual; this choice is misleading for
the subsidiary conditions (4) and (5) must be matched. This is possible only
fixing $\lambda=1$. Thus, the number of regulators must be at least two,
otherwise we can't get the coefficients $\lambda_i$ becoming arbitrarily
large.

So, let us take
\begin{equation}
c_{1}=\alpha -1
\hspace{0.3cm},
\hspace{0.3cm}
c_{2}=-\alpha
\hspace{0.3cm},
\hspace{0.3cm}
c_j=0\,\,\,;\,\,\,j>2\,\,,
\end{equation}
where the parameter $\alpha$ can assume any real value except zero and the
unity, so that condition (4) is satisfied. For $\lambda_1\,,\lambda_2
\rightarrow \infty\,\,,$
\begin{equation}
\Pi^{M}_{1}(k^2) \rightarrow \Pi^{(1)}(k^2)=
-\frac{e^2k^2}{(2\pi)}\int_{0}^{1}d\xi\,\,\xi\,(1-\xi)\frac{1}{{(M^2)}^
{\frac{1}{2}}}\,\,,
\end{equation}
where
\begin{equation}
M^2 \equiv m^2-\xi\,(1-\xi)k^2\,\,,
\end{equation}
and, consequently, $\Pi^{(1)}(0)=0$.

{}From (13) we find
\begin{eqnarray}
\Pi^{M}_{2}(k^2)=\frac{e^2}{4\pi m}\int_{0}^{1} d\xi\,\{\frac{m}{{[m^2
-\xi\,(1-\xi)k^2]}^{\frac{1}{2}}}+\frac{(\alpha -1)M_{1}}{{[M_{1}^2
-\xi\,(1-\xi)k^2]}^{\frac{1}{2}}} \nonumber \\
\,\,\,-\frac{\alpha M_{2}}{{[{M_{2}}^2
-\xi\,(1-\xi)k^2]}^{\frac{1}{2}}} \} \,\,\,.
\end{eqnarray}
Taking the limit $\lambda_1\,,\lambda_2 \rightarrow \infty$ for $k=0$, yields
\begin{eqnarray}
\Pi^{(2)}(0)=\frac{\alpha e^2}{4\pi m}(1-s)\,\,,\\
\nonumber \\
s \equiv sign(1-{\alpha}^{-1})\,\,.
\end{eqnarray}

For $0< \alpha < 1$, which corresponds to $s=-1$ and couplings $c_1$ and
$c_2$ having the same sign, $\Pi^{(2)}(0) \neq 0$; in this case the photon
acquires a topological mass, proportional to $\Pi^{(2)}(0)$, coming from
proper insertions of the antisymmetric sector of the vacuum polarization
tensor in the free photon propagator. If we assume that $\alpha$ is outside
this range, $s=1$, $c_1$ and $c_2$ have opposite signs and $\Pi^{(2)}(0)$
vanishes. We then conclude that this arbitrariness in the choice of the
parameter $\alpha$ reflects in different values for the photon mass. The
new parameter $s$ may be identified with the winding number of homotopically
nontrivial gauge transformations and also appears in lattice
regularization${}^{[6]}$.

Now we face another problem: which value of $\alpha$ leads to the correct
photon mass?
A glance at equation (13) and we realize that $\Pi^{(2)}(k^2)$ is ultraviolet
finite by na\"{\i}ve power counting. We were taught${}^{[7],[8]}$ that a
closed fermion loop must be regularized as a whole so to preserve gauge
invariance. However, having done that, we have affected the finite
antisymmetric piece of the vacuum polarization tensor and, consequently, the
photon mass. The same reasoning applies when, using Pauli-Villars
regularization, we calculate the anomalous magnetic moment of the
electron${}^{[7]}$; again, if care is not taken, we might arrive at a wrong
physical result.

In order to get rid of this trouble we should pick out the value of $\alpha$
that just cancels the contribution coming from the regulator fields. From
expression (21), we easily find that this occurs for $\alpha =1/2$, because
in this case the signs of the auxiliary masses are opposite, in account of
condition (5). We then obtain
\begin{equation}
\Pi^{(2)}(0)=\frac{e^2}{4\pi m}\,\,,
\end{equation}
in agreement with the other approaches already mentioned. We should remember
that Pauli Villars regularization violates parity symmetry in 2+1
dimensions${}^{[9]}$. Nevertheless, for this particular choice of $\alpha$,
this symmetry is restored as the regulator masses get larger and larger.

The result quoted above suggests that the ordinary parity-breaking
Pauli-Villars regularization, if carefully implemented, does not introduce
any resi\-dual contribution to the photon topological mass. In the causal
theory of the S-matrix${}^{[10]}$, this corresponds to the minimal splitting
of the causal distribution related to the vacuum polarization tensor in
$QED_3 {}^{[5]}$.

\newpage

{\large References}
\vspace{1.0cm}

\begin{sf}

\begin{description}
\item[{[1]}] S. Deser, R. Jackiw and S. Templeton, Ann. of Phys. {\bf 140}
(1982), 372;

\item[{[2]}] B. M. Pimentel, A. T. Suzuki and J. L. Tomazelli, Int. J. Mod.
Phys. {\bf A 7} (1992), 5307;

\item[{[3]}] R. Delbourgo and A. B. Waites, Phys. Lett. {\bf B 300} (1993),
241;

\item[{[4]}] B. M. Pimentel, A. T. Suzuki and J. L. Tomazelli, Int. J. Theor.
Phys. {\bf 33} (1994), 2199;

\item[{[5]}] G. Scharf, W. F. Wreszinski, B. M. Pimentel and J. L. Tomazell,
Ann. of Phys. {\bf 231} (1994), 185;

\item[{[6]}] A. Coste and M. L\"uscher, Nucl. Phys. {\bf 323} (1989), 631;

\item[{[7]}] W. Pauli and F. Villars, Rev. Mod Phys. {\bf 21} (1949), 434;

\item[{[8]}] P. Breitenlohner and H. Mitter, Nucl. Phys. {\bf B 7} (1968) 443;

\item[{[9]}] A. N. Redlich, Phys. Rev. {\bf D 29} (1984), 2366;

\item[{[10]}] G. Scharf, Finite Quantum Electrodynamics, Springer Verlag,
Berlin (1989).
\end{description}
\end{sf}
\end{document}